\documentclass[manuscript]{aastex}

\usepackage{xifthen}
\usepackage{color}
\usepackage{amsmath}
\usepackage{amssymb}
\usepackage{natbib}

\definecolor{Remove}{RGB}{255,0,0}
\definecolor{Replace}{RGB}{0,0,128}

\newcommand{\replace}[2]{{\textcolor{Remove}{\textbf{#1}}\textcolor{Replace}{\textbf{#2}}}}
\renewcommand{\replace}[2]{#2}

\newcommand{\chem}[3][]{%
\ifthenelse{\isempty{#1}}%
{{#2}\ensuremath{_{#3}}}%
{{#1}\ensuremath{_{#2}}{#3}}%
}

\newcommand{\usk}{\ensuremath{~}}
\newcommand{\kelvin}{\ensuremath{\usk\mathrm{K}}}
\newcommand{\microm}{\ensuremath{\usk\mathrm{\mu m}}}
\newcommand{\kilom}{\ensuremath{\usk\mathrm{km}}}

\newcommand{\gram}{\ensuremath{\usk\mathrm{g}}}
\newcommand{\gpcm}{\ensuremath{\usk\mathrm{g \usk cm^{-3}}}}
\newcommand{\mgpcm}{\ensuremath{\usk\mathrm{mg \usk cm^{-3}}}}
\newcommand{\cm}{\ensuremath{\usk\mathrm{cm}}}
\newcommand{\cmps}{\ensuremath{\usk\mathrm{cm \usk s^{-1}}}}
\newcommand{\cmpss}{\ensuremath{\usk\mathrm{cm \usk s^{-2}}}}
\newcommand{\cmsps}{\ensuremath{\usk\mathrm{cm^2 \usk s^{-1}}}}
\newcommand{\pc}{\ensuremath{\usk\mathrm{pc}}}
\newcommand{\pcc}{\ensuremath{\usk\mathrm{pc^{3}}}}
\newcommand{\ppcc}{\ensuremath{\usk\mathrm{pc^{-3}}}}

\begin{document}

\title{Atmospheric Habitable Zones in \replace{}{Cool} Y Dwarf Atmospheres}

\author{Jack S. Yates\altaffilmark{1,2}, Paul I. Palmer\altaffilmark{1,2},
Beth Biller\altaffilmark{3,2}, Charles S. Cockell\altaffilmark{4,2}
}
\email{j.s.yates@ed.ac.uk}

\altaffiltext{1}{School of GeoSciences, University of Edinburgh, UK}
\altaffiltext{2}{Centre for Exoplanet Science, University of Edinburgh, UK}
\altaffiltext{3}{SUPA, Institute for Astronomy, University of Edinburgh, Blackford Hill, Edinburgh, UK}
\altaffiltext{4}{UK Centre for Astrobiology, School of Physics and Astronomy, University of Edinburgh, UK}

\shorttitle{Atmospheric Habitable Zones} 
\shortauthors{Yates et al.}

\begin{abstract}
We use a simple organism lifecycle model to explore the viability of
an atmospheric habitable zone (AHZ), with temperatures that could
support Earth-centric life, which sits above an environment that does
not support life. \replace{We illustrate this 
idea using the object WISE J085510.83--0714442.5, which is a cool, free-floating brown 
dwarf. }{To illustrate our model we use a cool Y dwarf atmosphere, such as WISE 
J085510.83--0714442.5 whose 4.5--5.2 micron spectrum shows absorption features consistent with water vapour and clouds.} We allow organisms to adapt to their
atmospheric environment  (described by temperature, convection, and
gravity) by adopting  different growth strategies that maximize their
chance of survival  and proliferation. We assume a constant upward
vertical velocity  through the AHZ. We found that the organism growth
strategy is most  sensitive to the magnitude of the atmospheric
convection. Stronger  convection supports the evolution of more
massive organisms.  For a purely radiative environment we find that
evolved organisms have a mass that is an order of magnitude smaller
than terrestrial microbes, thereby defining a dynamical constraint on
the dimensions of life that an AHZ can support.  Based on a previously
defined statistical approach we infer that there are
of order $10^9$ \replace{}{cool} Y brown dwarfs in the
Milky Way, and likely a few tens of these objects are within ten
parsecs from Earth. Our work also has implications for exploring life
in the atmospheres of temperate gas giants.  Consideration of the
habitable volumes in planetary atmospheres significantly increases the
volume of habitable space in the galaxy.
\end{abstract}


\maketitle

\section{Introduction}

The recent discoveries of Earth-like planets orbiting their host
stars outside our solar system are beginning to challenge our
understanding of planetary formation and the development of
extra-terrestrial life. A common definition of whether a planet is
capable of supporting life is is whether the effective surface
temperature can sustain liquid water at its surface, which reflects
several factors including the evolution of the planet and star, and
the distance between them \citep{Kasting1993}.  Here, drawing on our knowledge of Earth
and inspired by previous theoretical work for the Jovian atmosphere we
argue that an atmosphere sitting above a potentially uninhabitable
planetary surface may be cool enough to sustain life.  By doing this
we define an atmospheric habitable zone (AHZ). 

The Earth's atmosphere contains a large number of aerosolized microbes
with concentrations ranging from $10^3~\mathrm{m}^{-3}$ to more than
$10^6~\mathrm{m}^{-3}$ of air, of which approximately 20\% are larger
than $0.5 \microm$ \citep{Bowers2012}.  The atmospheric residence time
of these organisms is highly uncertain but there is a growing body of
works that show that some organisms are metabolically active,
particularly in clouds
\citep{Lighthart1995,Lighthart1997,Fuzzi1997,Sattler2001,Cote2008,Womack2010,Gandolfi2013}.
Other solar system planets have been postulated to have a habitable
atmosphere. The Venusian surface temperature (${\sim}738$~K) is too
high to sustain liquid water, so based on Earth-centric definitions it
is uninhabitable.  At the cloud deck at ${\sim}55$~km, where
atmospheric temperatures are close to those at Earth's surface, liquid
water is more readily available and conditions are more amenable to
sustaining life \citep{Cockell1999,Schulze-Makuch2004,Dartnell2015}.  The Jovian
atmosphere has also been considered to be potentially habitable.
\citet{Sagan1976} described a microbial ecosystem that could optimize
a survival strategy to take advantage of their physical environment.

To illustrate the idea of the AHZ we focus on cool, free-floating
Y-class brown dwarfs \citep{Kirkpatrick2012}, thereby avoiding 
complications associated with any stellar effects on an atmosphere or on inhabiting organisms 
(e.g. radiation, stellar particles, and electromagnetic interactions). The 
coolest known object WISE J$085510.83-071442.5$ (henceforth W0855-0714)
has a mass $M_\mathrm{BD}$ of $6.5\pm3.5 M_\mathrm{Jup}$, a radius 
$R_\mathrm{BD}$ equal to $R_\mathrm{Jup}, 6.99 \times 10^{4} 
\kilom$, and an effective temperature $T_\mathrm{eff}$
of ${\sim}250 \kelvin$ \citep{Beamin2014,Faherty2014,Kopytova2014,Luhman2014}.
We expect that the upper atmosphere of cool objects similar to
WISE0855-0714 will have values for temperature and pressure similar to
Earth's lower atmosphere, and models and the latest spectra have
suggested that liquid  water in clouds may also be present
\citep{Faherty2014,Morley2014,Morley2014a,Skemer2016}.  Observed
spectra for cool brown dwarfs are consistent with significant dust
loading in the upper atmosphere \citep{Tsuji2005,Witte2011}. These
aerosols can provide charged surfaces on which prebiotic molecules,
necessary for life, could form \citep{Stark2014}. Prebiotic molecules
could also be delivered to the brown dwarf atmosphere via dust from
the interstellar medium \citep{MunozCaro2002}.  Based on current
understanding, M/L/T brown dwarf atmospheres also contain most of the
elements that are  thought to be necessary for life: C (in CH$_{4}$,
CO, CO$_{2}$), H (CH$_{4}$, H$_{2}$, H$_{2}$O, NH$_{3}$, NH$_{4}$SH),
N (N$_{2}$, NH$_{3}$), O (CO$_{2}$, CO, OH), and S (NH$_{4}$SH,
Na$_{2}$S) \citep[for example, see][]{Allard2012,Cushing2004,Cushing2006,Cushing2008,Cushing2011,Kirkpatrick2012}.

We develop the idea of a cool brown dwarf atmospheric sustaining life
in its atmosphere by using a simple 1-D model to describe the
evolution of a microbial ecosystem, following \cite{Sagan1976}, that
is subject to convection and gravitational
settling. The simplicity of our approach allows us to develop a
probabilistic understanding of the survival of individual organisms
under different environmental conditions.

In the next section we describe our numerical models. In 
\autoref{section:results} we present analytical and numerical results,
including a small number of sensitivity experiment that test our prior
assumptions. We discuss our results in a broader astrobiological
context and conclude the paper in \autoref{section:discussion}.

\section{Model Description}
\label{section:model}

We develop a simple atmospheric model that retains a sufficient level
of detail to describe the atmospheric environment that drives
variations in the lifecycle of the organism population. The organism
model draws from nutrient-phytoplankton
models used to described ocean biology on Earth (e.g.,
\cite{Franks2002}), but we allow organisms to determine an evolutionary growth
strategy that is best suited to the atmospheric environment.

\subsection{Brown Dwarf Atmosphere}
\label{W0855-0714model}

As described above, our illustrative calculations are based \replace{}{loosely} on the
object  W0855-0714 \citep{Luhman2014}. We are interested in the
region of the atmosphere that has temperatures in the range $258 \kelvin < T <
395 \kelvin$, which represent the lower and upper limits for life on
Earth \citep{McKay2014}. We define the AHZ as the atmospheric
region(s) that fall between those limits.

For our work, we define a $T-P$ profile based on the 200\kelvin, $\log
g = 5.0$ profile from the 1D model of \citet{Morley2014a},
assuming an atmosphere composed of 85\% \replace{H }{H$_2$} and 15\% He.
We assume these
gases exhibit near-ideal behaviour so we can calculate density using
the ideal gas law and altitudes can be calculated from scale heights. 
We define altitude at the bottom of the AHZ as $0 \kilom$, and
the P-T profile places upper edge of the AHZ at $\sim 105 \kilom$. We
calculate the luminosity of the object using the Stefan-Boltzmann law
and $T_\mathrm{eff}\sim 250K$ \citep{Luhman2014}.

\replace{}{
Our model atmosphere assumes the presence of liquid water in the AHZ
to support the biochemistry necessary to sustain life, as we know it
on Earth. This assumption restricts us to the coolest Y dwarfs and
also, for example, some cool gas giants. To illustrate our AHZ
hypothesis, we use a T--P (T$_{gas}$--P$_{gas}$) profile determined from a 1-D
hydrostatic model atmosphere simulation for Y dwarfs \citep{Morley2014}.
This model assumes equilibrium condensation processes such that
super saturation = 1. For a T$_{eff}$=200~K, $\log g$ = 5 object, the T--P
profile places the water phase transition between gas/liquid and ice
at a temperature lower than 273~K at 0.7 bar pressure \citep{Morley2014}.
To illustrate our model we use a cool Y dwarf atmosphere, for
example WISE 0855-0714 whose 4.5-—5.2 micron spectrum shows absorption
features consistent with water vapour and clouds \citep{Skemer2016}.
A supporting model calculation for this object (T$_{eff}$=250~K) shows that
the temperature profile intercepts the saturation vapour pressure at
$\simeq$273~K \citep{Skemer2016}, assuming equilibrium condensation
processes \citep{Morley2014}.  In nature, non-equilibrium processes
(super saturation $>$ 1) compete with equilibrium processes, allowing liquid water to exist at
super cooled (metastable) temperatures much lower than 273~K, e.g.
\cite{RogersYau1996, Helling2013, Helling2014}. }

\replace{}{Homogeneous freezing of pure liquid water is due to statistical
fluctuations of its molecular structure such that smaller drops ($<$5
microns) freeze spontaneously at temperatures closer to 243~K while
larger droplets freeze at slightly higher temperatures \citep{RogersYau1996};
similar empirical results are found for heavy water \citep{Wolk2001}.
On Earth, liquid water is not commonly found at such
low temperatures suggesting a role for heterogeneous freezing
processes. Liquid clouds are more commonly found at 253~K \citep{RogersYau1996}.
A cloud can be considered as a collection of independent
liquid droplets such that each droplet must be subjected to a
nucleation event before the whole cloud is frozen. A consequence of
this is that mixed-phase clouds are common over the coldest (polar)
geographical regions on Earth \citep{Morrison2012, Lawson2014, Loewe2016}.
Aerosols and aqueous solutions
influence nucleation of ice. Aerosol particles can act as ice
nuclei. Ice can form directly from the gas phase on suitable ice
nuclei via deposition and freezing heterogeneous nucleation processes
\citep{RogersYau1996}. Liquid water existing as a component of an
aqueous solution can significantly affect the temperature at which ice
begins to nucleate, depending on the water activity of the solution
\citep{Koop2000}.  Based on this we argue that homogeneous and
heterogeneous nucleation processes allow liquid water to exist at
temperatures much lower than 273 K. The lower limit for the AHZ
temperature is, however, determined by the coldest temperature 
($\sim 253~\mathrm{K}$) that can support Earth-based life.}

For simplicity, we use a constant convective vertical velocity $v_\mathrm{conv.}$
throughout the AHZ. We use 
values of $v_\mathrm{conv.}$ taken from a 3-D model of atmospheric dynamics 
\citep{Showman2013}, which was used to study L/T dwarfs. Because Y dwarfs are 
cooler we expect the associated convective velocities to be smaller. 
We use two values, $v_\mathrm{conv.} = 100\cmps$
and $v_\mathrm{conv.} = 1000\cmps$, which cover a range of plausible convection
scenarios,  to assess the effect of the windspeed on the final population
of organisms. In addition
we consider a radiative atmosphere, with $v_\mathrm{conv.} = 0.01\cmps$.

\subsection{Model of Organisms and their Lifecycle}

\begin{figure}[t]
\begin{center}
    \def\svgwidth{0.75\textwidth} \Large{
      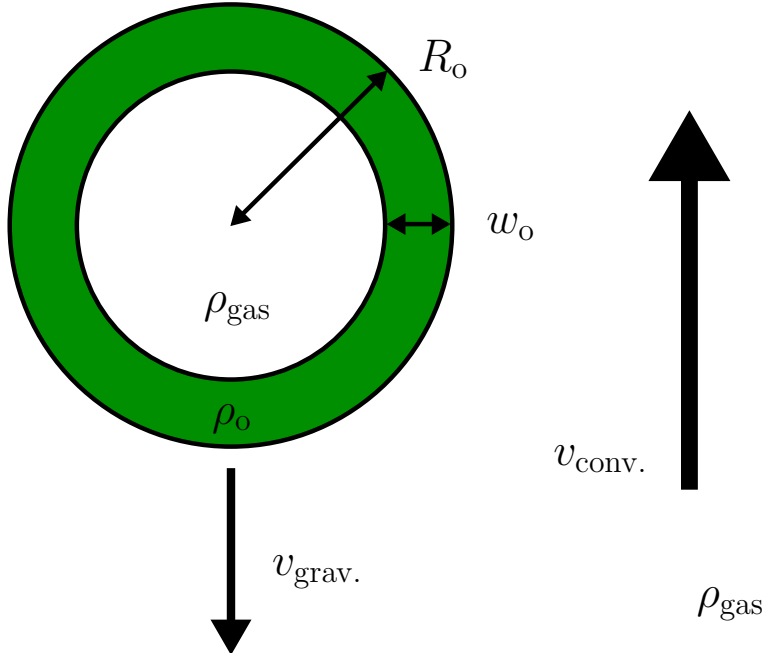 }
    \caption{Schematic of a model organism. The thickness
      and density of the organic skin layer are denoted by $w_0$ and
      $\rho_0$, respectively. The radius of the shell is denoted by
      $R_0$ and the density of the gas within the skin, in equilbrium
      with the atmosphere, is denoted by $\rho_\mathrm{gas}$.  The upward
      velocity is denoted by $v_\mathrm{conv.}$ and the terminal velocity due
      to gravity is denoted by $v_\mathrm{grav.}$.  }
    \label{figs:organism} 
\end{center}
\end{figure}

We describe an individual organism as a frictionless spherical shell, following
\cite{Sagan1976}. The shell is described by its radius, skin width,
mass, and density of the organic skin (\autoref{figs:organism}).
Organisms increase their mass by consuming biomass, described
below. Increasing an organism's mass increases its size and skin width
according to an organism-specific growth strategy, $G$, which is given
by
\begin{equation}
w_\mathrm{o} = (1 - G)R_\mathrm{o},
\end{equation}
where $w_\mathrm{o}$ is the organic skin thickness, and $R_\mathrm{o}$
is the radius of the sphere. Thus an individual organism with $G
\rightarrow 1$ is balloon-like, whilst organisms with $G \rightarrow
0$ are solid throughout.  We limit the skin densities
$\rho_\mathrm{o}$ to range between $0.5 \gpcm < \rho_\mathrm{o} < 1.5
\gpcm$. This is equivalent to densities greater than some light woods
and less than the density of heavy woods and bone. As a comparisons,
humans and microbes are approximately $1.0 \gpcm$, the density of
water, reflecting their bulk composition. For the purpose of this paper
we assume this skin is permeable so that the density of the organism
within the skin is the same at the atmosphere $\rho_o$. At the start
of each experiment organisms are uniformally distributed throughout
the AHZ.

Each organism has a lifecycle of growth, reproduction (subject to
sufficient growth), and death.  Organisms that are better suited to
their atmospheric environment will generally have more progeny, so
their parameter regime (analogous to genetic material) survives for
longer. At each timestep in our
model an organism eats, moves (dies immediately if they are now
outside the AHZ), reproduces subject to growth rate, and finally dies
if it is older than a specified half-life.

The growth of an organism is determined by its consumption of biomass.
Without any observational constraint we have been intentionally vague
on the composition of this biomass.  On Earth, organism growth is
typically limited by the availability of one element, compound, or
energy source at any one time. Populations of plankton in the Earth's
ocean, for example, are limited by the availability of trace elements
(such as P or N). We account for this by using a finite amount of
biomass that is available for consumption by the organisms. Consumed
biomass is returned to the atmosphere after an organism dies. The biomass is
initially distributed evenly throughout the AHZ, and after each
timestep any returned biomass is vertically distributed in the AHZ as a
function of the organism weight within each vertical layer.

Organisms move only by convection and gravitational settling 
(\autoref{figs:organism}). We assume laminar flow (defined by the Reynolds
number) so that the terminal velocity of the organism is given by
equating Stokes' drag force to the gravitation force:
\begin{equation}
v_\mathrm{grav.} = - \frac{\rho_\mathrm{o} -
  \rho_\mathrm{gas}}{\rho_\mathrm{gas}} \frac{g V_\mathrm{o}}{6 \pi
  \nu R_\mathrm{o}},
\label{eqn:stokes}
\end{equation}
where $v_\mathrm{grav.}$ is the sinking velocity of the organism
relative to the gas; $\nu$ is the kinematic viscosity; $g$ is the
gravitational acceleration ($16,670 \cmpss$); and $V_\mathrm{o}$,  the
volume of the organic material, is given by
\begin{equation}
V_\mathrm{o} = \frac{4 \pi \left(  R_\mathrm{o}^3 - \left(
  R_\mathrm{o} - w_\mathrm{o} \right)^3 \right)}{3} = \frac{4 \pi
  R_\mathrm{o}^3 \left(  1 -  G^3 \right)}{3}. 
\end{equation}
The vertical movement $\Delta h$ of the organism per timestep $\tau$
is given by $(v_\mathrm{conv.} + v_\mathrm{grav.})\tau$ so that the
vertical position $h$ after $n$ timesteps $h_{n} = h_{n - 1} + \Delta
h$. On Jupiter and Saturn the different observed bands are thought to
be sites of upwelling and downwelling. Models have suggested that
Jupiter's banded structure is stable, and have also suggested that
similar structures might be common in brown dwarf atmospheres
\citep{Showman2013}. Based on this study, we assume that convection is
stable such that at some latitudes the upward velocity will be
approximately constant, and at these latitudes the body might sustain
bands of life.

Each organism has a half-life of 30 Earth days, which is reasonable
for Earth microbes. We retain organisms that meet the following
criterion: $2^{-\frac{\tau}{T_{1/2}}} > \mathcal{U}\left[0, 1
  \right]$, where $\mathcal{U}\left[a, b \right]$ is a number drawn
from the uniform distribution with limits $a < b$, and all other
variables are as previously defined. We acknowledge that some
microbes are very short-lived or can spend many years
cryogenically frozen before being revived \citep{Gilichinsky2008}. For
simplicity, we assume that an organism dies if it moves outside the AHZ, but we discuss this further in \autoref{section:organismdiscussion}. 

Each organism attempts to reproduce at every timestep. The number of
progeny, $n_c$, is determined by dividing the mass of the organism by
its ``reproduction mass'', rounding down and subtracting one (the
organism retains some mass after it reproduces). If $n_c \ge 1$, we
split the organism into $n_c$ progeny (and itself), each with a
slightly perturbed set of inherited characteristic to account for
genetic mutation. Therefore, as an example, if the organism mass is $3.1
m_{\mathrm{repr.}}$ it will have two progeny each of mass $m \approx
m_{\mathrm{repr.}}$. The reproduction mass is close to the birth mass
of the organism, with some small variation.  Inherited characteristics
vary according to a distribution with the mean being the value of the 
parameter for the parent. Growth strategies use
a normal distribution, limited to 0.01--0.99, with a standard
deviation of 0.05; densities use a normal distribution, limited to
$0.5$--$1.5 \gpcm{}$, with a standard deviation of $0.05 \gpcm$;
reproduction masses are varied according to a log-normal
distribution with a standard deviation of 10\% of the mean. The
initial skin width and size can be inferred from the density, strategy 
and the initial mass. 

\section{Results}
\label{section:results}

\subsection{Analytical Model Estimates}

We estimate organism sizes and masses for a given convective windspeed
using the model described above, assuming a zero net vertical velocity
so that an organism can float indefinitely in the convective updraft. 

As described above, we limit the densities of the organisms to $0.5
\gpcm < \rho_\mathrm{o} < 1.5 \gpcm$, while gas densities in the AHZ
range from $0.4 \mgpcm$ to $1.2 \mgpcm$. We can then assume
$\left(\rho_\mathrm{o} - \rho_\mathrm{gas} \right) / \rho_\mathrm{gas}
\approx \rho_\mathrm{o} / \rho_\mathrm{gas}$ so that
\autoref{eqn:stokes} becomes:
\begin{equation}
R^2_\mathrm{o} \rho_\mathrm{o} = \frac{9 v_\mathrm{conv.} \nu
  \rho_\mathrm{gas}}{ 2 g }.
\label{eqn:analytical}
\end{equation}
We assume that the gravitational acceleration is effectively constant
throughout the AHZ ($g = 16,670 \cmpss$); and that $\nu$ ranges from
$0.1 \cmsps$ at the hottest part to $0.2 \cmsps$ at the coldest part
of the AHZ, with corresponding values of $\rho_\mathrm{gas}$ of $1.2
\mgpcm$ and $0.4 \mgpcm$.  We account for the inner cavity by
absorbing the change in mass into the value of $\rho_\mathrm{o}$ to
determine an ``effective density'', $\rho_\mathrm{eff.}$. For a value
of $G \approx 0.5$, $\rho_\mathrm{eff.} \approx 7 \rho_\mathrm{o} / 8$
assuming $\rho_\mathrm{gas} \ll \rho_\mathrm{o}$. The growth strategy
therefore does not make a significant difference until $G > 0.8$,
after which $\rho_\mathrm{eff.} < \rho_\mathrm{o} / 2$. Most organisms
are in the regime where $\rho_\mathrm{eff.} \approx \rho_\mathrm{o}$. 

\begin{figure}[p]
\centering \includegraphics[width=\textwidth]{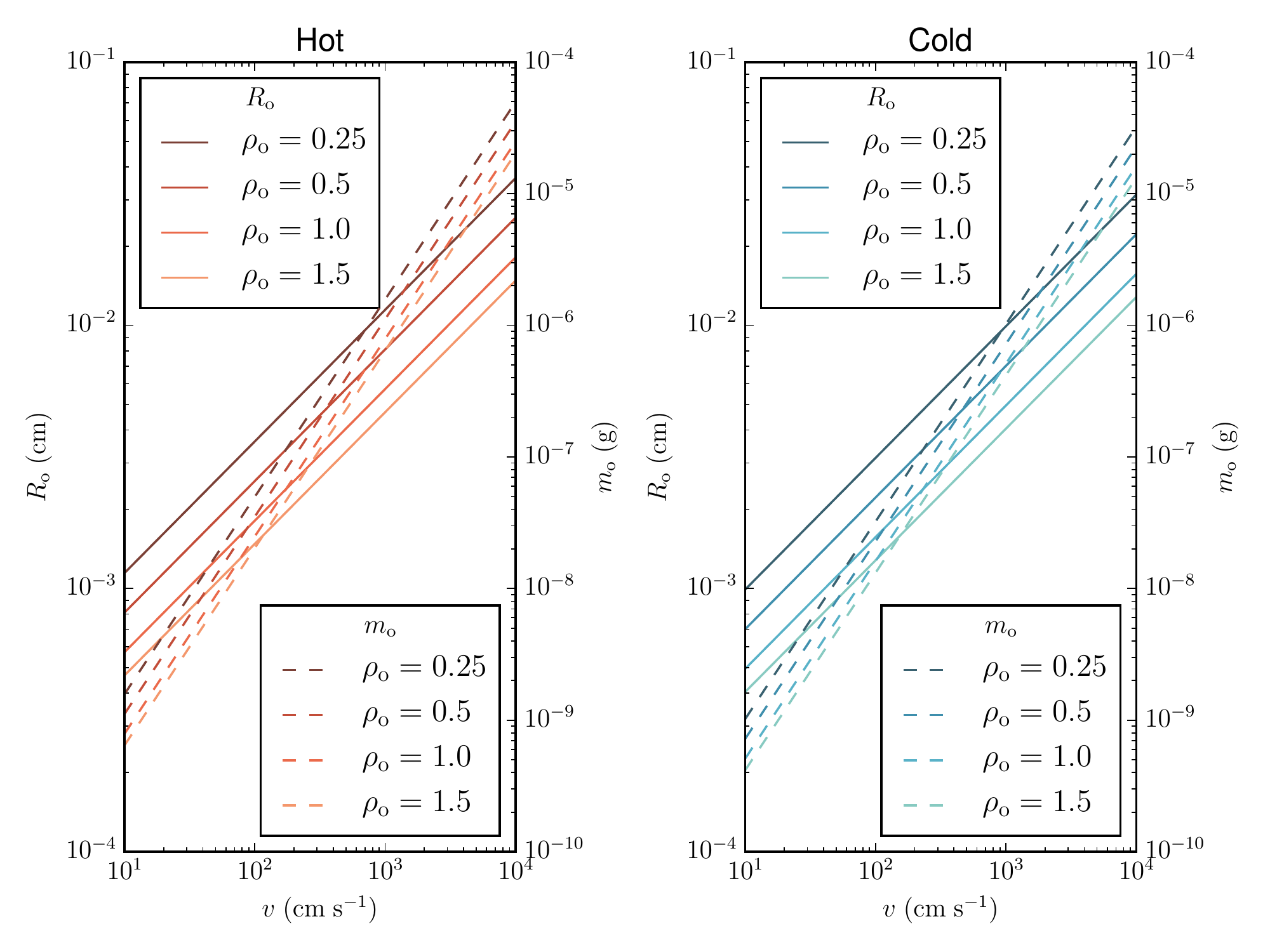}
\caption{\label{figs:analytical} Expected size and mass of an organism
  for a given windspeed determined using the analytical model.  The
  line colour denotes the density of the organism  including the
  cavity.  Solid lines denote the radius, $R_\mathrm{o}$, required for
  the sinking rate to match the upward windspeed. The dashed lines
  denote the corresponding mass, $m_\mathrm{o}$. Hot denotes the
  hotter, lower AHZ ($395 \kelvin$) and cold denotes the colder, higher AHZ ($258 \kelvin$), which use corresponding values for $\nu$ and $\rho_\mathrm{gas}$ ($0.1 \cmsps$ and $1.2 \gpcm$, or $0.2 \cmsps$ and $0.4 \gpcm$, respectively).}
\end{figure}

Based on these assumptions, \autoref{figs:analytical} show typical
values for the vertical position, size, and effective density of
organisms for a given windspeed. We find that for moderate windspeeds
($< 1000 \cmps$) in the convective zone, a typical organism should be
a few orders of magnitude more massive and about a factor of ten
larger than a terrestrial microbe (${\sim}10^{-12} \gram, <10^{-4}
\cm$). These estimates will be a useful check on the full model
results. We find that different values of $\nu$ and
$\rho_\mathrm{gas}$ indicative of the cold and hot limits of the AHZ
change masses by less than a factor of 2 and sizes even less.

\subsection{Numerical Model Estimates}

Our control model experiment has a convective windspeed of $100\cmps$,
a timestep of six hours, a initial population of 100 organism with an
approximate mass of $10^{-9}\gram$ distributed randomly (with a
uniform distribution) across the AHZ. Each organism is initialized
with random properties. We run an ensemble of 20 simulations, each for
100 Earth years.

We test each simulations for steady state conditions by looking at the
stationarity of the total number of organisms; a trend or changing
variance would indicate that the population was still undergoing
changes and had not yet settled to a viable strategy. To achieve this
we use an augmented Dickey-Fuller (henceforth ADF) test \citep{Said1984} on the last 75 years of
population data and found that all runs had reached steady state to a
very high significance ($p < 0.01$). We present results that represent the mean
model state from the last Earth year of each experiment.

\begin{figure}[p]
\centering
\includegraphics[width=\textwidth]{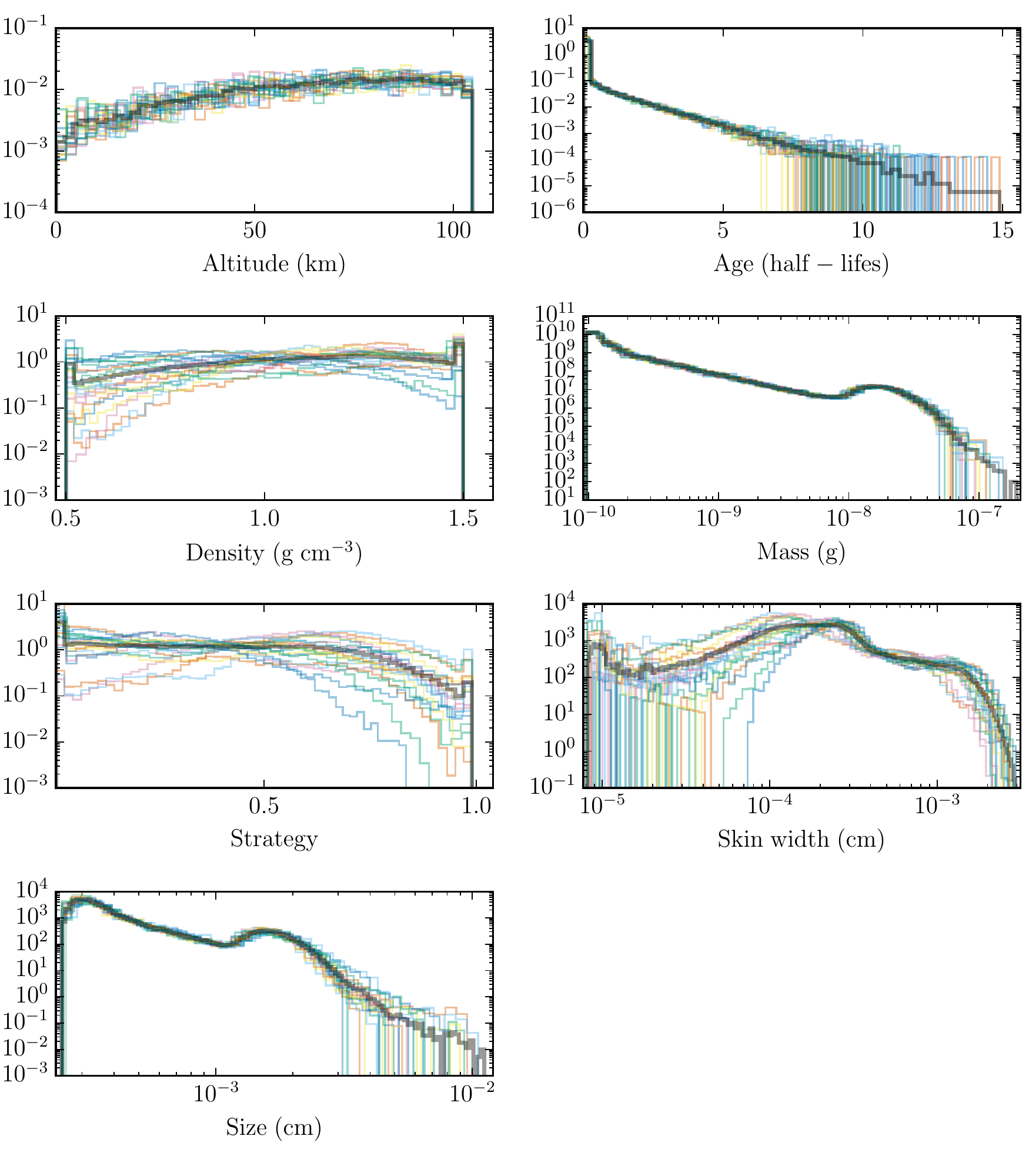}
\caption{Normalized frequency distributions of organisms in the AHZ as a
  function of altitude, age (as half-lives), density, mass, growth
  strategy, skin width, and size. The dark lines denote the average of
  the ensemble of the 20 individual model run shown by coloured
  lines. The frequency distribution is normalized such that its integrates
  to unity between the lowest and highest values.
  \label{figs:default-hist}}
\end{figure}

\autoref{figs:default-hist} shows that organisms are approximately
evenly spread throughout the AHZ with a small skew towards to the
top. An approximate even distribution
suggest that the organism have found a mass/size strategy to support a
stable population.  In general this strategy is found within a few
years.  The age distribution of the organisms follow the expected
half-life distribution well, with the exception of a large number of
very young organisms. These small organisms are also visible in the
mass, skin width and size distributions. We find this population is a
persistent feature of our experiments, but individuals have a short
residence time as they are rapidly convected out of the AHZ. The
population is an artifact of our reproduction scheme: individually
they are unviable but are frequently produced as they only require a
small amount of mass.  The densities are evenly spread across the
allowed range with a small skew to higher densities, suggesting that
there is most likely no significant effect on the dynamical behaviour
of the organisms. Densities are forced within the range $0.5 \gpcm <
\rho_\mathrm{o} < 1.5 \gpcm$, which accounts for the small excesses at
either end of the distribution.  Aside from the population of small,
short-lived organisms, the mass distribution peaks at around $2 \times
10^{-8} \gram$, with most organisms being between $10^{-9}$ and
$10^{-7} \gram$, which is consistent with the analytical model. 
Organisms within this range are relatively stable
in the convection, with residence times of 30 days or more.

We find that the growth strategy favoured by the organisms in the
control calculation is skewed towards lower values of $G$,
i.e. particles that are more solid. As discussed above for the
analytical calculations, the effect of the growth strategy on the
organism's size or effective density goes approximately with the cube
of $G$. Thus values of $G < 0.8$ have very little effect on the
dynamical behaviour of the organism, which we see in the distribution.
The distribution of skin widths and size is as expected with the peak
value for the skin width between $0.3$ and $3 \times 10^{-3} \cm$,
consistent with the analytical model. 

\subsection{Sensitivity Runs} 
\label{section:sensitivity-runs}

\begin{deluxetable}{lcccccc}

\tablewidth{0pt}
\tablecaption{Sensitivity Runs\label{table:sensitivity}}
\tablehead{Description & $m_\mathrm{input}~(\mathrm{g})$ & $B$ &
$\tau~(\mathrm{hrs})$ & $v_\mathrm{conv.}~(\mathrm{cm \usk s^{-1}})$ &
$T_{1/2}~(\mathrm{days})$}

\startdata
Control run		& $10^{-9}$ 		& 1 & 6 & 100  & 30 \\ 
High windspeeds		& $10^{-8}$ 		& 1 & 2 & 1000 & 30 \\ 
Radiative windspeeds	& $10^{-13}$		& 1 & 6 & 0.01 & 30 \\ 
Population increase	& $10^{-9}$ 		& 3 & 6 & 100  & 30 \\ 
Initial input mass	& $2\times10^{-9}$	& 1 & 6 & 100  & 30 \\ 
Short half-life		& $10^{-9}$		& 1 & 6 & 100  & 15 \\ 
Long half-life		& $10^{-9}$		& 1 & 6 & 100  & 60 \\ 
\enddata

\tablecomments{Initial conditions for each set of sensitivity runs. Shown is the mean initial organism mass $m_\mathrm{input}$, the approximate biomass factor $B$ relative to the control run, the timestep in Earth hours $\tau$, the windspeed $v_\mathrm{conv.}$, and the half-life of the organism $T_{1/2}$. } 

\end{deluxetable}


\newcommand{\rcb}{radiative-convective boundary}

We run a small set of sensitivity runs to test our prior model
assumptions; for each sensitivity experiment we run a ensemble of 10
replicates. \autoref{table:sensitivity} summarizes the We use the ADF to ensure
that all resulting populations were stable and sustainable. We find
that our results are not significantly sensitive to changes in the
initial conditions for the organisms (e.g. amount of available
biomass) and these results are not discussed further.

We run the models with vertical velocities of $1000 \cmps$ and $0.01 \cmps$.  The
slower of these velocities is intended to replicate windspeeds in the
radiative zone. Most models put the \rcb{} somewhere above the AHZ,
but it is possible for it to be below the AHZ or even inside
it. Changing the wind speed will affect the range of masses that could
be sustained in the AHZ: generally, slower convection supports the
evolution of lighter organisms and higher convection supports the
evolution of heavier organisms.

There are two implications of changing the vertical velocity that we
consider when choosing the lower and upper values.  First, varying the
windspeed will change the distribution of organism sizes and masses
that can sustain a population. To address this we also change the
range of masses that are allowed within the model based on the
analytical model results described above.  Second, changing the
vertical velocity changes the distance over which organisms can move
during one timestep. We have addressed this by co-adjusting the
timestep so that the Courant Friedrichs Lewy criterion is met \citep{Courant1928}. We
find that our results are not significantly affected by changing the
model timestep.

\begin{figure}[p]
\centering
\includegraphics[width=\textwidth]{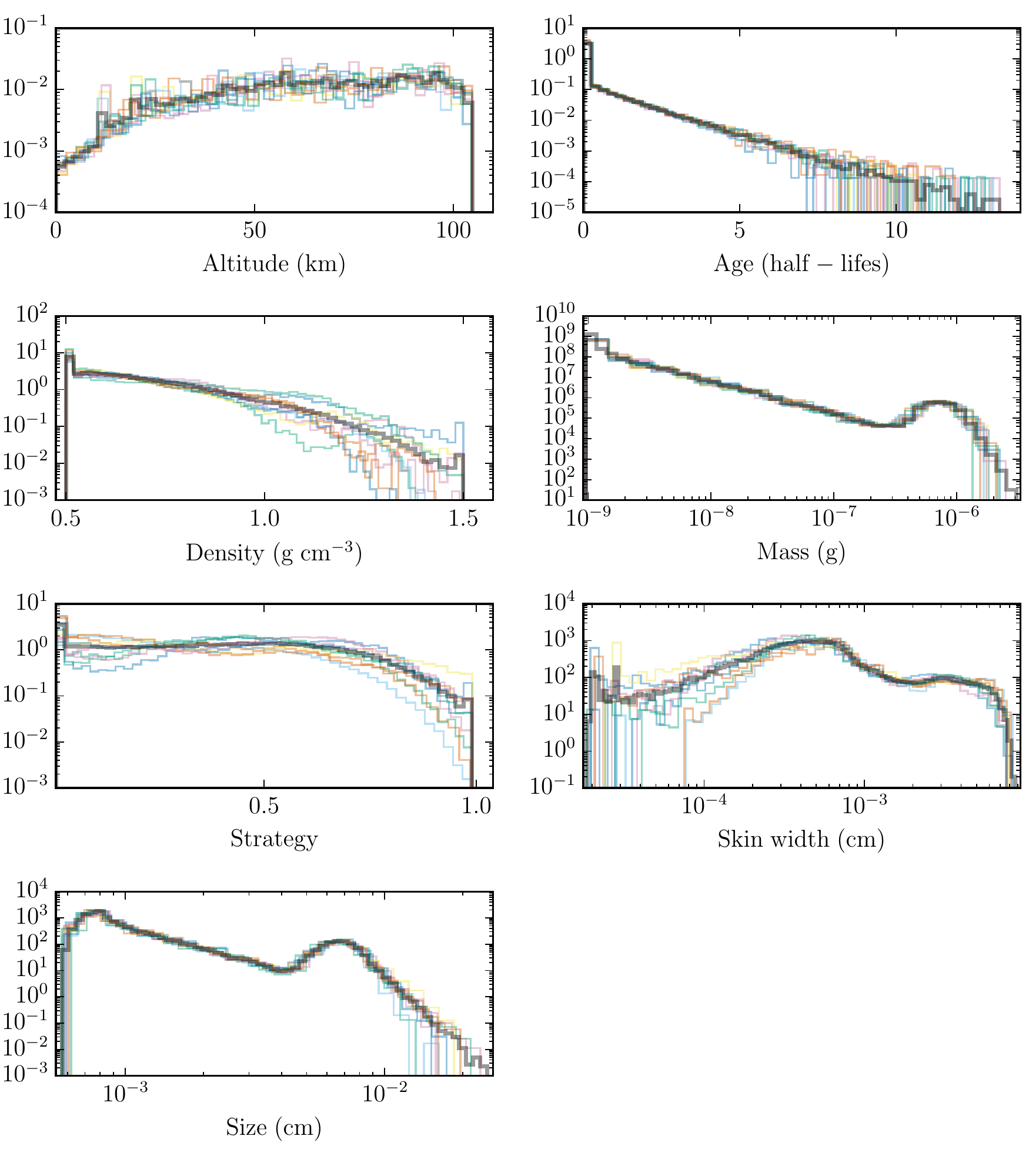}
\caption{\label{figs:810-hist}
  As \autoref{figs:default-hist} but using a convective velocity of $1000 \cmps$.}
\end{figure}

\begin{figure}[p]
\centering
\includegraphics[width=\textwidth]{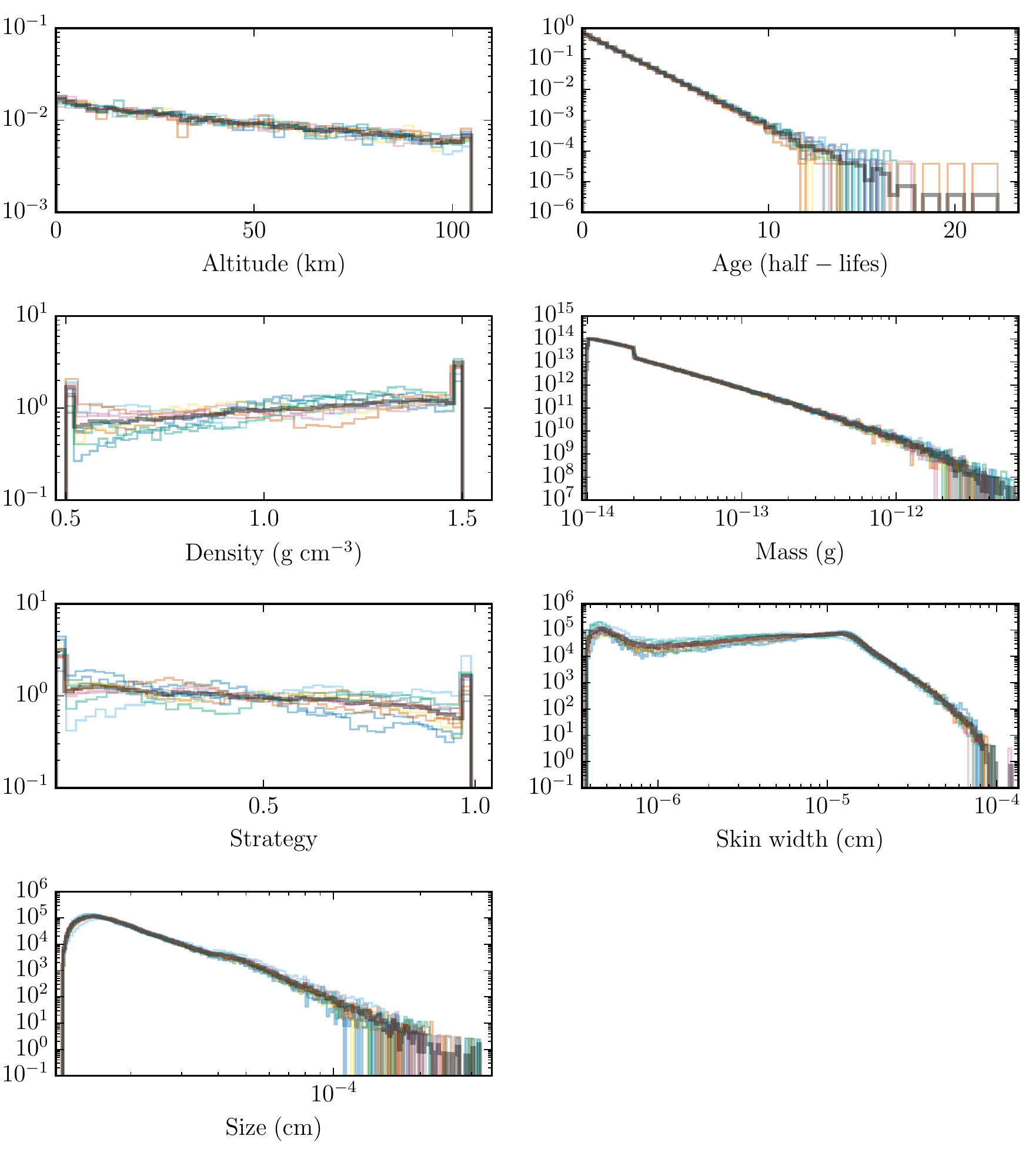}
\caption{\label{figs:600-hist}
  As \autoref{figs:default-hist} but using a radiative vertical velocity of $0.01 \cmps$.}
\end{figure}

\autoref{figs:810-hist} and \autoref{figs:600-hist} shows that the
organism population distributions corresponding to vertical velocities
of $1000 \cmps$ and $0.01 \cmps$, respectively, are different to those
from the control run. We find that skin widths peak at around $5 \times 10^{-4} \cm$ and $3 \times 10^{-3} \cm$ with sizes showing two peaks at $7.5 \times 10^{-4} \cm$ and $8 \times 10^{-3} \cm$. Thus in most cases organisms have small cavities, which is also reflected in the strategy graph. Masses drop off steadily from $10^{-9} \gram$ to a second peak at $10^{-6} \gram$. Here the analytical calculations have somewhat overestimated the required mass to sustain a population. Age and altitude graphs are consistent with the control run. 


For the radiative experiment, the organism distributions are generally
similar to those of the control experiment.  This is a skew in the
vertical distributon of the organisms, where there are approximately
twice as many organisms at the bottom of the atmosphere when compared
with the top. The age distribution is close to a perfect half-life
distribution. This non-convective environment also appears to support
a slightly wider range of growth strategies.  Unlike the control
experiment, the non-convective environment cannot support the highest
masses with the distribution falling off near-monotonically in
log-space. This is consistent with our analytic model that shows that
for every tenfold decrease in windspeed organisms should decrease in
mass by a factor of $10^{3/2}$.  The expected mass of approximately
$10^{-14} \gram$ is comparable to the mass of a terrestrial virus
\citep{Johnson2006}.  Terrestrial microbes are a factor of ${\sim}100$
too massive to be supported by this environment. This environment
could be considered as habitable if life could be described by something
much smaller than a terrestrial microbe.

%


The half-life of organisms will impact the ability of the AHZ to
sustain life. For example, shorter organism half-lives mean greater
turnover of biomass and a reduced emphasis on the atmospheric
transport in determining observed variations. To address this we used
a half-life that was half and double that of the 30-day control run
value. For organisms with a 15-day half-life there is an excess of
very young organisms compared with old organisms, which is due to a
higher turnover of biomass that increases birth rates.  Other
distributions are not significantly different from the control case.

\section{Discussion}
\label{section:discussion}

Here, we put our model results into a wider context. We also discuss
briefly the associated implications for habitability.

\subsection{Cool Brown Dwarf Spatial Frequency and Galactic Significance}

\newcommand{\dmax}{\ensuremath{d_\mathrm{max}}}
\newcommand{\Vmax}{\ensuremath{V_\mathrm{max}}}
\newcommand{\VVmax}{\ensuremath{V/V_\mathrm{max}}}

There are only tens of known cool brown dwarfs of which
WISE 0855-0714 is the coolest. These objects are inherently 
faint and consequently are difficult to detect unless they are 
nearby. To understand how common AHZs might be, we extrapolate
from the number of known Y dwarf objects (though the results are 
likely applicable to other types of bodies).

To estimate the frequency of these cool brown dwarfs, we determine
spatial densities based on the distances of known objects, following
\citet{Kirkpatrick2012}.  For objects that are uniformly distributed
in space, with some objects yet to be discovered, we perform a
\VVmax{} test to assess the completeness of a sample.  First, we
calculate the interior volume $V_i$ of each real object, i.e. the
spherical volume centred on Earth, with radius $d_i$, where $d_i$ is
the distance to object $i$. The sample is assumed to be complete out
to some distance \dmax{}. Within the corresponding interior volume,
\Vmax{}, the mean value of $V_i/\Vmax{}$ should be 0.5. 
Using the value of \dmax{}, we can estimate the spatial density of the objects
by dividing the number of objects closer than \dmax{} by \Vmax{}.
Here, we calculate \dmax{} for each spectral type, rather than define
a value of \dmax{} and assess the completeness \citep{Kirkpatrick2012}. 
(Naturally, we reject objects with $V_{i}/\Vmax{} > 1$. In this case 
we remove the objects and recalculate \Vmax{} until all objects have
$V_{i}/\Vmax{} > 1$.) 

We analyzed brown dwarfs of spectral type Y0 or later with published
parallaxes, using the data from \citet{Tinney2014} and references
therein. Where there is more than one published parallax, we take a
mean weighted by the inverse of the square of the uncertainties. We 
group objects into two spectral type categories: spectral types 
$\mathrm{Y}0-0.5$ and $\geq\mathrm{Y}1$. 


For the $\mathrm{Y}0-0.5$ category we include 9 objects and reject 4. 
We find $\dmax{} = 9.6 \pc$, $\Vmax{} = 3.6 \times 10^{3} \pcc$ and a 
spatial density of $\sim 2.5 \times 10^{-3} \ppcc$. For the 
$\geq\mathrm{Y}1$ objects, we include 3 objects and reject 1. Here we 
find $\dmax{} = 10.85$, $\Vmax{} = 5.4 \times 10^{3} \pcc$ and a spatial
density of  $\sim 0.6 \times 10^{-3} \ppcc$. 

Assuming the Milky Way is a disk with a diameter of
$4 \times 10^{4} \pc$ and a thickness of $600 \pc$ (corresponding to
a volume of $\simeq$750$\times 10^9 \pcc$), we can project these numbers 
on to the galaxy as a whole. These calculations produce very approximate 
numbers of $2 \times 10^{9}$ objects with spectral type $\mathrm{Y}0-0.5$
and $0.5 \times 10^{9}$ objects with spectral type $\geq\mathrm{Y}1$ 
in the galaxy. Further, we expect that there should be around 
10 $\mathrm{Y}0-0.5$ dwarfs and 2 $\geq\mathrm{Y}1$ dwarfs within 10 pc of 
Earth. 

Clearly these samples are highly incomplete so we consider this a
conservative estimate. Brown dwarf cool as they age on timescales of
billions of years \citep{Baraffe2003}. Hotter objects will eventually
develop an AHZ, while for cooler objects the AHZ will descend in the
atmosphere and will contract as the effective temperature falls and
the lapse rate grows. As the AHZ descends the associated pressure will
increase so that the nature of the organisms as we describe them here
will undoubtedly change.

\subsection{Implications for Habitability}
\label{section:habitability}

Estimating the total potential biomass achievable in an aerial
biosphere of a brown dwarf for which we have very limited knowledge of
the elemental composition of the atmosphere, let alone the nutritional
requirements of a hypothetical biota, cannot be done with accuracy. We
still have much to learn about how nutrient limitation and
co-limitation influences the distribution of biomass on
Earth. However, from a more general point of view, we would expect
that the upper limit of biomass theoretically achievable in a brown
dwarf atmosphere to be determined by limitation of specific
nutrients. Once more data are available on the atmospheric composition
of very cool brown dwarf atmospheres it may even be possible to predict which
elements would limit a potential biota in these
environments. We are not aware of any modelling studies predicting the presence of phosphorus 
in significant quantities.

The implications of the AHZ concept for habitability in the galaxy are
significant. In the most simplistic view there are, conservatively,
billions of cool brown dwarfs in the galaxy and hundreds within a few
tens of parsec of the Earth. Some of these will be targets for
characterisation for next-generation telescopes in less than a decade,
although their inherent faintness makes them difficult to find in
surveys.

When searching for habitable environments, we naturally take an
Earth-centric focus on terrestrial planets that receive their energy
from the host star. \replace{In the case of W0855-0714 it is clear that
self-heating is sufficient to produce an atmospheric environment that
can support liquid water. }{Thermal spectra of W0855-0714 shows features consistent
with atmospheric water vapour and clouds \citep{Skemer2016}, suggesting 
that self-heating is sufficient to produce an atmosphere with liquid 
water at habitable temperatures}. This observation is not limited to
application to brown dwarfs; Jupiter receives approximately as much
heating from the Sun as from its core, as does Saturn. Gas giants in
other stellar systems could also potentially support similar
biomes. Our work provides further evidence for the habitability of
planets such as Venus that have an uninhabitable surface. Whilst the
dynamics of Venus' atmosphere will likely be very different to that of
a cool brown dwarf or gas giant, it seems likely that with some
\replace{tweaking }{adjustment }of the properties one could model an organism that could
sustain a population in the Venusian AHZ indefinitely. Thus we support
the idea that the inner edge of the circumstellar habitable zone is
not a hard limit on habitability. Detecting this aerial biosphere with
current and next-generation telescopes will depend on the range and
resolution of the spectrally-resolving instruments, and also the range
and magnitude of byproduct gases that the organisms produce.

\subsection{Reflections on our Model Organism}
\label{section:organismdiscussion}

We described organisms as individual frictionless hollow spheres with
a permeable skin. Organisms that successively evolved within the
prescribed physical environment of the AHZ eventually shaped the final
cohort that were characterized by their radius and skin
thickness. While our organism model does exist in nature, e.g., pollen
spore with air sacs, we did not impose any further constraint that
could impact their atmospheric lifetime in the AHZ.

We did not consider the coalescence of similar organisms or
deposition onto existing airborne particles. Earth bacteria
can occur as agglomerations of cells or attach themselves to airborne
particulate matter, such as pollen, or aqueous-phase aerosol
\citep{Jones2004}. In our simple approach, we can consider an organism as a
single entity or an agglomeration of many entities. Heavier organisms sink 
in the atmosphere at greater speeds. Organisms made of lots of individual
entities could employ an additional survival strategy: organisms could attach to each other or to atmospheric particulates as temperature decreases (as they ascend), and disperse as temperature increases
(as they descend), allowing them to self-correct to find the centre of the AHZ. 
This strategy is similar to that proposed by \citet{Sagan1976}, where organisms increase in mass and split into low-mass daughter cells as they sink to the lower parts of the AHZ.

We also did not consider cryogenic freezing of organisms above the top of the AHZ. It is conceivable that an organism could pass through the upper boundary of the AHZ, spend some time in statis and subsequently sink back into the AHZ, whereupon it would thaw and reactivate. We anticipate that low-mass or low-density organisms would have greater survivability in this scenario.

Surface roughness is another parameter that organisms could use to 
adapt to their environment. A rough surface would increase
the drag properties of the organism, resulting in slower movement that
deviates significantly from Stoke's formula \citep{Md2015}.  Recent
empirical evidence also suggests that if a microorganism has the
ability to maintain a charge it could substantially decrease its drag
and speed up its trajectory \citep{Md2015}. If the charge could be
manipulated it could be used to stabilize the altitude of organisms.

In nature, there a number of examples where animals manipulate their
body drag through water.  Fish can alter the drag between their skin
and medium by altering their body smoothness by excreting
high-molecular weight polymer compounds and surfactants
\citep{Daniel1981} or in the case of sharks by altering their body
geometry \citep{Dean2010}.  Studies have also proposed that seals and
penguins actively use bubble-mediated drag reduction, required to launch
themselves out of water \citep{Davenport2011}.

While our model of an organism is rudimentary there do exist in nature a
number of examples in which animals control their movement in a
laminar flow. Therefore it is possible that our microorganisms could
have evolved to stabilize their movement without the need for
necessarily changing their physical dimensions.

\section{Conclusions}
\label{section:conclusions}

We used a simple organism lifecycle model to explore the viability of
an atmospheric habitable zone (AHZ), with temperatures that could
support Earth-centric life. The AHZ sits above some uninhabitable environment, 
such as an uninhabitable surface (e.g. as on Venus) or a hot dense atmosphere 
(e.g. the lower atmosphere of a brown dwarf or gas giant).
We based our organism model on previous work that
explored whether the Jovian atmosphere could support life.
We illustrated this idea using a cool Y brown dwarf, for example object W0855-0714.

\replace{}{
Our atmospheric model assumed availability of liquid water that is necessary
to support the biochemistry associated with life. There exist valid
counter arguments for the presence of liquid water in the AHZ of a
cool brown dwarf (e.g., changes in metallicity may affect partial
pressure of water, although metallicity has to change by orders of
magnitude to affect H$_2$O \citep{Helling2008}), which may suggest that
our model is valid only for a small subset of available cool Y dwarfs (and
a range of gas giants). The formation history of brown dwarfs result
in a wide range of atmospheric chemical composition environments, e.g.
\cite{Madhusudhan2016}. In the authors' view the most
compelling argument for the presence of liquid water is the observed thermal
spectra of WISE 0855-0714 that showed evidence of atmospheric water
and clouds \citep{Skemer2016}. Based on our understanding of Earth's
atmosphere, water clouds cannot exist without the presence of liquid
water.}
We also assumed a constant upward vertical velocity through the
AHZ and model organisms that float in the convective updrafts. Our modelled organisms
can adapt to their atmospheric environment by adopting different
growth strategies that maximize their chance of survival and producing
progeny. We found that the organism growth strategy is most sensitive to
the magnitude of the atmospheric convection. Stronger convective winds
support the evolution of more massive organisms with higher gravitational 
sinking rates, counteracting the upward force, while weaker
convective winds results in organisms that need less mass to overcome
the upward convective force. For a purely radiative environment we found that the
successful organisms will have a mass that is ten times smaller than
terrestrial microbes, thereby putting some dynamical constraints on
the dimensions of life that the AHZ can support. 

We explored the galactic implications of our results by considering
the likely number of Y brown dwarfs in the galaxy, based on the number
we know. We calculate that of order $10^9$ of these objects reside in
the Milky Way with a few tens within ten parsecs of
Earth. Some of these close objects will be visible to large telescopes
in the next decade. Our work has focused on brown dwarfs but it also
has implications for exploring life on gas giants in the solar system
and exoplanets, which have uninhabitable surfaces. \replace{If we account for
the habitable volumes residing in planetary atmospheres we can
significantly increase the volume of habitable space in the galaxy.}{Our calculations suggest a significant upward revision of the volume of habitable space in the galaxy.}

\acknowledgements
We thank the anonymous referee for thoughtful and
thorough comments on the manuscript. \replace{}{We gratefully acknowledge
discussions about the presence (or not) of liquid water in cool Y
dwarf atmospheres with Franck Selsis, J\'{e}r\'{e}my Leconte, Christiane Helling,
Tim Garrett, and Caroline Morley.} J.S.Y. was supported by the
U.K. Natural Environment Research Council (Grant NE/L002558/1)
through the University of Edinburgh's E3 Doctoral Training
Partnership. P.I.P. gratefully acknowledges his Royal Society
Wolfson Research Merit Award. BB gratefully acknowledges support
from STFC grant ST/M001229/1.

\bibliographystyle{apj.bst}
\bibliography{library.bib}

\end{document}